\documentclass[twocolumn,superscriptaddress,amsmath,amssymb,aps,longbibliography,prb]{revtex4-2}
\usepackage[colorlinks=true,urlcolor=blue,citecolor=blue,linkcolor=blue]{hyperref}
\usepackage{txfonts}
\usepackage{graphicx}
\usepackage{dsfont}
\usepackage{bbm}
\usepackage{bm}
\usepackage{booktabs}
\usepackage{amsmath}
\usepackage{algorithm}
\usepackage{algorithmicx}
\usepackage{algpseudocode} 
\usepackage{graphicx}
\usepackage{dcolumn}
\usepackage{rotating}
\usepackage{longtable, booktabs, threeparttablex}

\newcommand{\Eq}[1]{Eq.~(\ref{#1})}

\begin{document}

\title{Deep variational free energy prediction of dense hydrogen solid at 1200K}
\author{Xinyang Dong}
\affiliation{Beijing National Laboratory for Condensed Matter Physics and Institute of Physics, \\Chinese Academy of Sciences, Beijing 100190, China}
\affiliation{AI for Science Institute, Beijing 100080, China}

\author{Hao Xie}
\affiliation{Beijing National Laboratory for Condensed Matter Physics and Institute of Physics, \\Chinese Academy of Sciences, Beijing 100190, China}
\affiliation{Department of Astrophysics, University of Zürich, Winterthurerstrasse 190, 8057 Zürich, Switzerland}

\author{Yixiao Chen}
\affiliation{Princeton University, Princeton, NJ 08544, USA}

\author{Wenshuo Liang}
\affiliation{DP Technology, Beijing 100080, China}

\author{ Linfeng Zhang}
\affiliation{AI for Science Institute, Beijing 100080, China}
\affiliation{DP Technology, Beijing 100080, China}

\author{Lei Wang}
\email{wanglei@iphy.ac.cn}
\affiliation{Beijing National Laboratory for Condensed Matter Physics and Institute of Physics, \\Chinese Academy of Sciences, Beijing 100190, China}

\author{Han Wang}
\email{wang\_han@iapcm.ac.cn}
\affiliation{Laboratory of Computational Physics, Institute of Applied Physics and Computational Mathematics, Fenghao East Road 2, Beijing 100094, China}

\begin{abstract}
We perform deep variational free energy calculations to investigate the dense hydrogen system at 1200 K and high pressures. In this computational framework, neural networks are used to model the free energy through the proton Boltzmann distribution and the electron wavefunction. By directly minimizing the free energy, our results reveal the emergence of a crystalline order associated with the center of mass of hydrogen molecules at approximately 180 GPa. This transition from atomic liquid to a molecular solid is marked by discontinuities in both the pressure and thermal entropy. Additionally, we discuss the broader implications and limitations of these findings in the context of recent studies of dense hydrogen under similar conditions.

\end{abstract}

\maketitle

\section{Introduction}
Dense hydrogen exhibits a diverse range of fascinating physical phenomena~\cite{RevModPhys.66.671} despite being composed of the simplest element. At low temperatures, the system adopts a variety of solid-state phases, which have been observed in both experiments~\cite{mao1988synchrotron, loubeyre1996x, PhysRevB.82.060101, ji2019ultrahigh} and simulations~\cite{PhysRevB.101.094104, li2024high}. 
At high temperatures, the system exists in a liquid phase, consisting of either molecular or atomic hydrogen~\cite{wigner1935possibility, landau1943relation}. Understanding the transition between these two liquid states is essential for both planetary modeling and high-pressure physics~\cite{weir1996metallization, fortov2007phase, knudson2015direct, zaghoo2017conductivity, celliers2018insulator}.

In spite of extensive computational studies of the liquid-liquid transition in the hydrogen system, its underlying physics remains unclear. This is partly due to that traditional methods, such as density functional theory (DFT) or ab-initio molecular dynamics, are limited by their accuracy~\cite{lorenzen2010first, hinz2020fully, bund2021isotope}. 
More accurate methods like quantum Monte Carlo (QMC), while providing higher precision, also face challenges in achieving unbiased sampling of proton degrees of freedom~\cite{attaccalite2008stable, morales2010evidence, pierleoni2016liquid, PhysRevLett.76.1240, Mazzola2014}.
Moreover, most of these approaches suffer from high computational costs, which limit simulations to relatively small systems or short timescales.
Recent efforts based on machine-learned force fields~\cite{cheng2020evidence} have enabled simulations at much larger sizes and longer timescales. However, the possible trade-off in accuracy during the supervised training~\cite{karasiev2021liquid} may limit their capacity to fully capture the complex behavior of the underlying physical processes.
For example, the latest molecular dynamics studies based on the machine learning force field trained with QMC data~\cite{tirelli2022high,istas2024liquidliquidphasetransitionhydrogen} find first-order atomic liquid to molecular liquid transition around 1200K, which is qualitatively different from the supercriticality behavior reported in Ref.~\onlinecite{cheng2020evidence}. 
In this context, it is noteworthy to mention recent findings on the solid-state phase around 1200 K~\cite{niu2023stable, goswami2024high} obtained using machine learning force fields trained with QMC data, which sharply contrast with all of previous calculations and experimental observations. 
The discrepancies between these various experimental and computational studies might be caused by multiple reasons.
On the experimental side, due to the difficulty of performing measurements under high pressures, experiments don't directly measure melting but instead detect anomalies in the Raman signal.
Additionally, the temperature at which melting occurs is indirectly inferred.
On the computational side, recent findings in Refs.~\onlinecite{niu2023stable, goswami2024high} indicate that the accuracy of the potential energy surface can qualitatively affect the predicted phases, highlighting the importance of incorporating precise many-body energies beyond the DFT level.
These considerations motivate a careful and independent re-examination of dense hydrogen in the debated parameter regime.

In order to investigate possible phases of dense hydrogen in this parameter regime, we perform a variational free energy calculation~\cite{xie2023deep} to study the system at 1200 K. This computational method combines flow-based generative models and neural network-based wavefunctions to model the variational density matrix of the dense hydrogen system.
By modeling the proton Boltzmann distribution and electron wavefunction using two deep neural networks and minimizing the free energy, we obtain an approximate solution for the system's equation of state at the QMC level of accuracy, which is approximately 1 milli-Hartree per atom.
Additionally, we incorporate physical knowledge into the model through a pretraining strategy based on proton ensembles and utilize a machine learning force field as the base distribution of the flow model.
This computational approach is closely related to coupled electron-ion Monte Carlo~\cite{Pierleoni2004} and Langevin dynamics~\cite{Attaccalite2008a} methods. The key difference is that we replace the Monte Carlo sampling of protons with a variational calculation, which provides direct access to the free energy and entropy of the system.
All of these methods take into account many-electron correlations~\cite{PhysRevB.88.014115} and thermal effects, but exhibit different characteristics when applied to finite-temperature structure prediction~\cite{kruglov2023crystal}. 
Molecular dynamics or Monte Carlo simulations are unbiased methods for sampling the potential energy surface. 
However, they may face challenges in capturing transitions between different phases, which correspond to different modes in the probability distribution, thus hindering the discovery of novel structures.
In contrast, the variational free energy calculation has the ``mode seeking'' behavior~\cite{Goodfellow-et-al-2016}, which can lead to convergence at local minima in the free energy landscape. Fortunately, having access to the variational free energy provides a way to check the optimality of the obtained solution. 

This paper is organized as follows: Section~\ref{sec:method} introduces the computational framework of the deep variational free energy approach, which contains a flow model enhanced by machine learning force fields and neural network wavefunction. Section~\ref{sec:results} presents benchmark results and analyzes the discovered solid-state phases in the dense hydrogen system. Finally, Section~\ref{sec:discussion} discusses the implications and limitations of the present study.

\section{Methods}
\label{sec:method}

\subsection{Variational free energy framework}

We consider a system of $N$ hydrogen atoms confined in a periodic simulation cell, with cell vectors $\bm a = (\bm a_1, \bm a_2, \bm a_3)$ and reciprocal vectors $\bm a^\ast = (\bm a_1^\ast, \bm a_2^\ast, \bm a_3^\ast)$.
In our convention, $\bm a_\mu \cdot \bm a_\nu^\ast = \delta_{\mu\nu}$, where $\delta$ is the Kronecker delta. 
Throughout this work, the coordinates of the particles and the cell vectors are scaled by a dimensionless parameter $1/r_s$, which represents the length scale of the system. 
For a cubic simulation cell with side length $L$, $r_s=\left(3/4 \pi N\right)^{1/3}L/a_B$, where $a_B$ is the Bohr radius.
The proton and electron coordinates are denoted by $\bm{X} = \{ \bm{x}_I \}$ and $ \bm{R} = \{ \bm{r}_i \}$, respectively.
For convenience, we sometimes treat the protons and electrons on equal footing and introduce $\bm \Xi = \{\bm\xi_\alpha\} = \bm R \bigcup \bm X$ as a unified notation for the coordinates of both electrons and protons.
Using this convention, the Hamiltonian of the system is written as
\begin{align}
  H = 
    -\frac1{2 r_s^2}\sum_{i} \nabla_{\bm r_i}^2 + 
    \frac 1{2r_s}\sum_{\alpha\beta\bm n}^\ast \frac{q_\alpha q_\beta}{\vert \bm \xi_\alpha - \bm \xi_\beta + \bm n \vert } \, ,
\end{align}
where $q_\alpha$ is the charge of particle $\alpha$,
and $\bm n = n_1\bm a_1 + n_2\bm a_2 + n_3\bm a_3$, with $n_\mu\in \mathbb Z$ and $\mu=1,2,3$.
The ``$\ast$'' in the summation indicates that if $\bm n=0$, the terms with $\alpha= \beta$ are omitted.
We will focus on spin unpolarized systems with $N/2$ spin-up and $N/2$ spin-down electrons in this work.

\begin{figure}[]
  \centering
  \includegraphics[trim={1cm 1cm 1cm 1cm}, clip, width=\columnwidth]{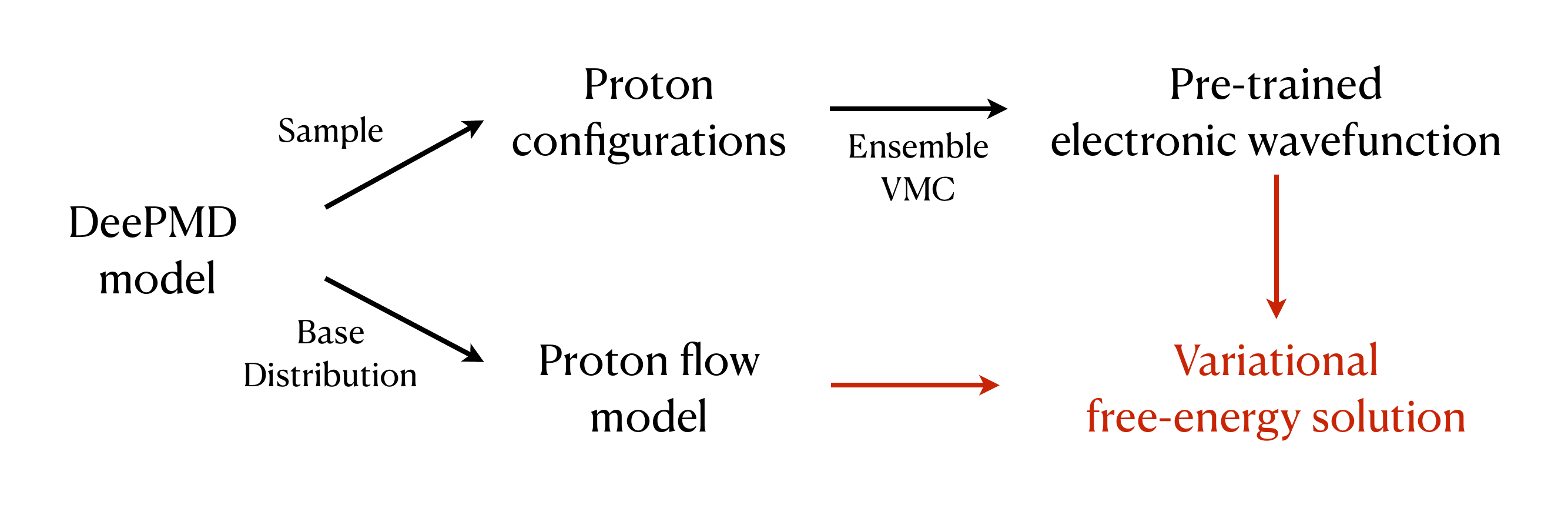}
  \caption{The flowchart of pretraining and finetuning of the hydrogen variational density matrix. A force field provides both the proton configurations for pretraining of the wavefunction and also the base distribution of the proton flow model. Variational free energy calculation finetunes the networks and provides a solution for the dense hydrogen system.}
  \label{fig:workflow}
\end{figure}

We aim to study the equation of state of the hydrogen system near the melting line~\cite{niu2023stable, goswami2024high}, where protons are treated as classical point particles, and electrons are assumed to stay in the instantaneous ground state for any given proton configuration.
To achieve this goal, we employ the variational free energy approach~\cite{PhysRevLett.121.260601,PhysRevLett.122.080602,xie2023deep}. The corresponding objective function is given by 
\begin{align}
F = \mathop{\mathbb{E}}_{\boldsymbol{X}\sim p(\boldsymbol{X})} [k_B T \ln p(\boldsymbol{X}) + E(\boldsymbol{X})] \, , 
\label{eq:free_energy}
\end{align}
with $p(\mathbf{X})$ the proton Boltzmann distribution and $E(\boldsymbol{X})$ the internal energy of the system for given proton coordinates $\boldsymbol{X}$.

In the variational free energy method, the proton Boltzmann distribution is parameterized by a normalizing flow network~\cite{xie2023deep}
\begin{equation}
p( \boldsymbol{X}) 
=  p_\text{base}(\boldsymbol{Z}) 
 \left| \det{ \left( \frac{\partial \boldsymbol{Z}}{\partial \boldsymbol{X}} \right) } \right|
=  \frac{e^{-E_\mathrm{base}(\boldsymbol{Z})/ k_BT }}{\mathcal{Z}_\mathrm{base}} 
 \left| \det{ \left( \frac{\partial \boldsymbol{Z}}{\partial \boldsymbol{X}} \right) } \right| \, , 
 \label{eq:flow}
\end{equation}
where $E_\mathrm{base}$ is the potential energy surface (PES) provided by the base model,
$\boldsymbol{Z}$ are a set of collective variables that are transformed from the proton coordinates $\boldsymbol{X}$ via a neural network with a non-degenerate Jacobian.
By parametrizing the proton Boltzmann distribution in this way, the normalizing flow model learns to correct the potential bias in the base PES model.
The Jacobian determinant of the transformation in Eq.~(\ref{eq:flow}) ensures that the normalization factor of the proton distribution remains independent of the flow model parameters.
With a perfectly optimized flow model, the probability distribution will converge to the target distribution $p(\boldsymbol{X}) \propto e^{-E(\boldsymbol{X})/k_B T}$.

To avoid systematic bias and maintain a fully variational calculation framework, we compute the PES by parametrizing a trial wavefunction with neural networks,
similar to {the variational Monte Carlo (VMC) method. {The ground-state energy is defined as}
\begin{align}
  E(\boldsymbol{X}) = \mathop{\mathbb{E}}_{\boldsymbol{R} \sim |\Psi(\bm R; \bm X)|^2 }
  \left[
  \frac{H \Psi(\bm R; \bm X) }
  {\Psi(\bm R; \bm X)}
  \right] \, ,
  \label{eq:elecwfn}
\end{align}
{where \(\Psi(\bm{R}; \bm{X})\) denotes the ground-state electronic wavefunction for a given proton configuration \(\bm{X}\).
It is noteworthy that we use an ensemble VMC approach, as described by Xie et al.~\cite{xie2023deep}, in which the wavefunction is universally parameterized and jointly optimized for any proton coordinates sampled from the given distribution to achieve the ground state.}
Physical quantities such as energy, entropy, and pressure can be estimated from the samples generated by the proton Boltzmann distribution and electronic wavefunction after the free energy optimization.

During the variational free energy training process, the objective function in \Eq{eq:free_energy} is optimized by adjusting the parameters of both the flow model [\Eq{eq:flow}] and the electron wavefunction network [\Eq{eq:elecwfn}] via stochastic optimization.
At each optimization step, we use Monte Carlo algorithm to draw samples of proton and electron coordinates from both models via ancestral sampling.
To better leverage prior knowledge about the hydrogen system within the optimization framework, we design a two-stage optimization scheme as shown in Figure~\ref{fig:workflow}.
The key idea is to train a machine-learning-based potential energy model—DeePMD \cite{zhang2018deep}—and use it to facilitate the optimization of both the flow model and the electron wavefunction (see Appendix~\ref{appsec:deepmd} for details of the DeePMD model).
More specifically, in the pretraining stage, we optimize the electronic wavefunction on an ensemble of proton configurations~\cite{Gao2021a, Scherbela2021} generated through molecular dynamics sampling using DeePMD. This make sure the electron wavefunction is at a good starting point before the joint optimization. Next, the wavefunction and the flow model—where the DeePMD model serves as the base distribution in \Eq{eq:flow}—are jointly optimized during the variational training process. 
This approach ensures that the variational free energy calculation serves as a fine-tuning stage, which corrects possible biases in the DeePMD model for the proton distribution while preserving a fully variational framework and yielding a proton distribution consistent with QMC-level accuracy.

In the following subsection, we will discuss general considerations for designing neural network architectures for variational free energy optimization of liquid hydrogen.

\subsection{Neural networks}

As shown in Eq.~\eqref{eq:flow}, the proton Boltzmann distribution is constructed using a flow model.
A physically valid proton Boltzmann probability must be invariant under the translation of all protons, periodic transformations of individual protons, and permutations of protons. To ensure these symmetries, the flow model must itself be invariant under translations and periodic transformations, while also being equivariant under proton permutations.
In practice, we achieve this by using a neural network that maps the proton coordinates $\boldsymbol{X}$ to a set of collective variables $\boldsymbol{Z}$ through an equivariant backflow transformation inspired by a Ferminet-like design~\cite{pfau2020ab}. {For a comprehensive explanation of the flow network, readers are directed to Appendix~\ref{appsec:flow}.

For the electron wavefunction ansatz, we design a modified Ferminet~\cite{pfau2020ab} / Deepsolid~\cite{li2022ab} network architecture to satisfy the required symmetry constraints in the liquid phase.
The translation invariance of the electron wavefunction under translations of both the electrons and protons requires
$
  \Psi(\bm R; \bm X) = \Psi(\bm R + \bm s; \bm X + \bm s)
$, where $\bm s$ is any constant vector in $\mathbb{R}^3$.
Additionally, we impose the {special} twisted boundary conditions (TBC)~\cite{lin2001twist} on the electronic degrees of freedom to mitigate finite-size effects. Under these conditions, wrapping an electron around the simulation cell introduces an additive phase change to the wavefunction, while the wavefunction remains periodic with respect to the proton degrees of freedom
\begin{align}
    &\Psi( \dots, \bm r_i + \bm n, \dots; \bm X) 
    = 
    e^{2\pi i \bm w\cdot \bm n}\Psi( \dots, \bm r_i, \dots; \bm X),\\
    &\Psi( \bm R; \cdots, \bm x_I + \bm n, \cdots) 
    = 
    \Psi( \bm R; \cdots, \bm x_I, \cdots),
\end{align}
with $\bm n = n_1 \bm a_1 + n_2 \bm a_2 + n_3 \bm a_3$, $n_\mu \in \mathbb{Z}$ and $\mu = 1, 2, 3$, and $\bm w$ the specified twist of the wavefunction.
Furthermore, the wavefunction should be symmetric (anti-symmetric) with respect to the permutation of protons (electrons of the same spin) $\sigma$
\begin{align}
  &\Psi(\bm R; \sigma(\bm X)) = \Psi(\bm R; \bm X) \, , \\
  &\Psi(\sigma(\bm R); \bm X) = (-1)^{\vert \sigma\vert}\Psi(\bm R; \bm X) \, .
\end{align}

The two major improvements we made compared to the network architectures in Ref.~\onlinecite{xie2023deep} are as follows:
First, we introduce a multi-scale transformation for constructing periodic features, which enhances the feature representation in both the flow and electron wavefunction neural networks.
Detailed explanations can be found in Appendix Section~\ref{appsec:feat}.
Second, we incorporate a gated attention layer within the message-passing process of the electron wavefunction ansatz.
This design improves the network's generalization ability, as elaborated in Appendix~\ref{appsec:elecwfn}.

\section{Results}
\label{sec:results}
In this section, we first present benchmark results to demonstrate the accuracy of the electron wavefunction, then we move on to physical results.

\subsection{Benchmarks}

\begin{table}[]
  \centering
  \caption{MAE {in ground-state energy} of ensemble VMC training for systems containing 32 hydrogen atoms.}
\label{tab:mae-32H}
\begin{tabular}{@{}cccccc}
  \multicolumn{6}{@{}r@{}}{\footnotesize{mHa/atom}} \\
  \toprule
   && \phantom{123}1000K & \phantom{123}1200K & \phantom{123}1400K & \\ [0.5ex] 
  \midrule
    &$r_s = 1.40$ & \phantom{123}0.8(1) & \phantom{123}1.2(5) & \phantom{123}0.8(2) & \\
  \midrule
  &$r_s = 1.44$ & \phantom{123}1.6(3) & \phantom{123}1.6(3) & \phantom{123}1.6(3) & \\
  \bottomrule
\end{tabular}
\end{table}


The variational free energy calculations at different pressures utilize a neural network wavefunction with the same architecture. 
Therefore, we aim to demonstrate the flexibility of the electron wavefunction in this section, so that the calculation will not be biased towards one particular phase during the free energy optimization procedure.

First, to demonstrate the accuracy of our electron wavefunction neural network in single-configuration variational calculations, we benchmark it against the reference data for 64 hydrogen atoms from Ref.~\cite{PhysRevLett.120.025701}. As shown in Appendix~\ref{appsec:single-conf}, our wavefunction consistently achieves lower ground-state energies than the reference data across all thermodynamic states in single-configuration training tasks, including both molecular and atomic configurations.

Next, we benchmark the accuracy of the multi-configuration-trained wavefunction at the pretraining stage to test its transferability and ensure it is a good starting point for the variational free energy training.
In the pretraining step, the wavefunction is pretrained using \textit{ensemble} {VMC} training.
Proton configurations are sampled using DeePMD simulations over a grid in the temperature-$r_s$ thermodynamic state space. Temperatures range from 1000K to 2000K in increments of 200K, while $r_s$ ranges from 1.35 to 1.50 in steps of 0.02.
A total of 6400 configurations are generated, from which a random subset of 512 configurations is used at each training step. The batch size for electron configurations is set to 8 for each proton configuration.
To test the transferability of our wavefunction, we benchmark its performance by comparing the {ground-state} energies predicted by the multi-configuration trained model against those of the single-configuration trained model. Table~\ref{tab:mae-32H} reports the mean absolute error (MAE) calculated using three test configurations for $r_s = 1.40$ and $r_s = 1.44$ at three different temperatures. These tested thermodynamic states are not included in the training ensemble. In all cases, the MAE remains below 2~mHa per atom, demonstrating that our pre-trained wavefunction is able to reliably provides accurate energy estimates across an ensemble of configurations, thus satisfying the requirements for variational free energy training.

\begin{figure}[tbh]
  \centering
  \includegraphics[scale=1]{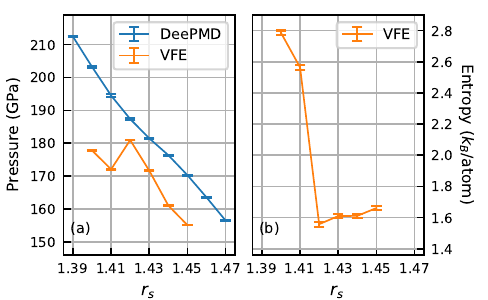}
  \caption{The (a) pressure and (b) entropy per particle versus density parameter computed at $T=1200$K for $N=32$ hydrogen atoms. The variational free energy calculations reveal significant features due to the formation of a crystalline structure at low densities.}
  \label{fig:pressure}
\end{figure}

\begin{figure*}[tbh]
  \centering
  \includegraphics[scale=1]{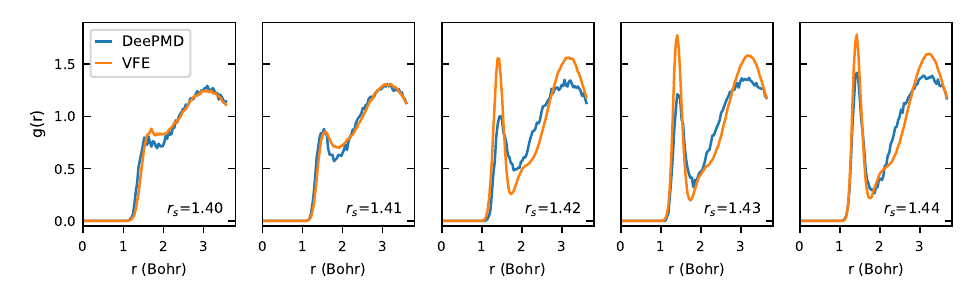}
  \caption{The proton-proton radial distribution function of the DeePMD base and the deep variational free energy prediction computed at $T=1200$K.}
  \label{fig:rdf}
\end{figure*}

\begin{figure}[tbh]
  \centering
  \includegraphics[width=\columnwidth]{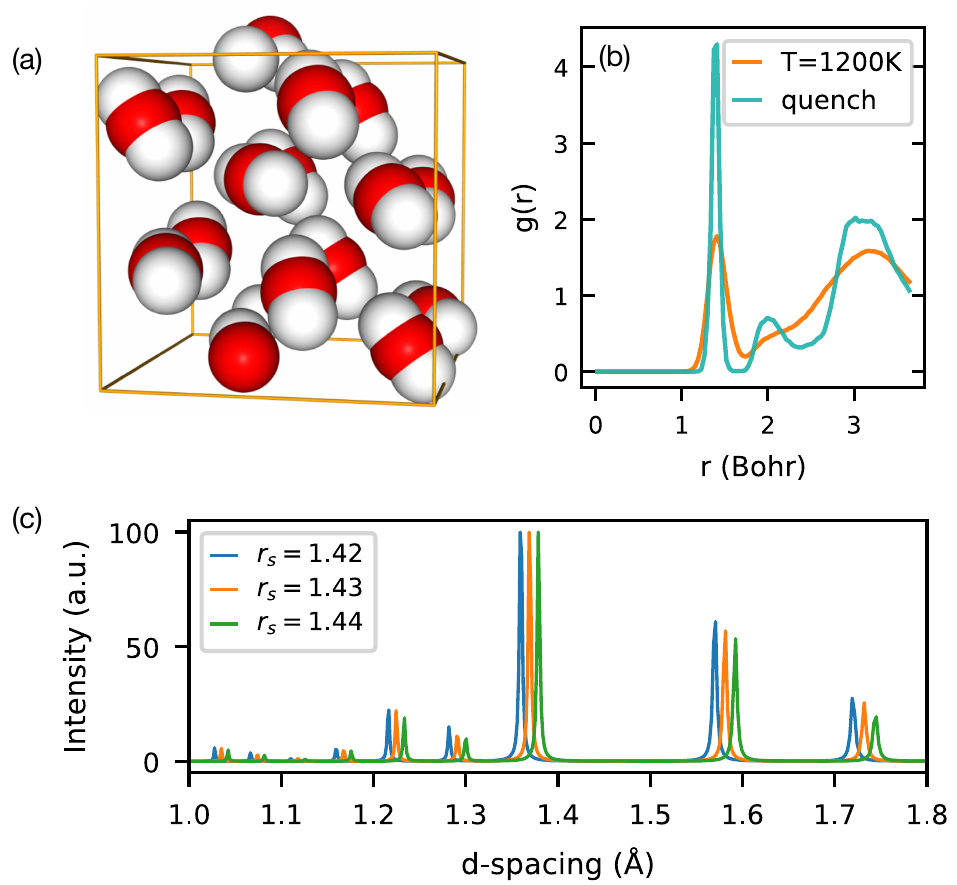}
  \caption{(a) The radial distribution function of protons computed at $T=1200$K and $r_s=1.43$ along with the one after quench.  (b) A snapshot of the protons shown as white spheres. The red spheres denote the center of mass of hydrogen molecules. (c) The simulated XRD spectra for an ensemble of configurations obtained at 1200K.}
  \label{fig:com}
\end{figure}

\subsection{High-temperature solids}
In this subsection, we present the results of variational free energy optimization for the dense hydrogen system at 1200K, with a focus on the emergence of crystalline order. Both the flow model and the electron wavefunction are trained within the variational free energy framework. At each optimization step, 512 proton configurations are sampled from the trained Boltzmann distribution, and 16 electron configurations are sampled for each proton configuration.

Figure~\ref{fig:pressure} shows the pressure and entropy as a function of the density parameter $r_s$ at $T=1200$K. The change in slope at $r_s=1.42$ in the pressure curve indicates a discontinuous phase transition.
Since the partition function of the base distribution enters the entropy calculation (see Eq.~\eqref{eq:flow}), we performed annealed importance sampling~\cite{neal2001annealed} on the DeePMD model to compute its free energy $-\ln \mathcal{Z}_{DP}$ in order to obtain the entropy of the system.
The entropy curves shown in panel (b) provides further insight into the two states, with the sharp decrease in entropy at $r_s=1.42$ suggesting a transition to a more ordered state.

To further analyze the phase transition, we plot the proton-proton radial distribution function (RDF) of the trained model alongside the DeePMD base distribution in Figure~\ref{fig:rdf}. Compared to the uniform base distribution employed in Ref.~\onlinecite{xie2023deep}, the DeePMD model provides a more informative reference for the proton distribution. 
However, around the critical point at $r_s = 1.42$, the variational free energy optimization leads to noticeable deviations in the radial distribution function (RDF).
The transition into a molecular state is characterized by the emergence of the molecular peak at around $1.4 ~ a_B$ as the $r_s$ value increases. Additionally, the presence of extra features around $2 ~ a_B$ and $3 ~ a_B$ suggests that the system may develop additional structures other than a molecular liquid.
The convergence of the flow-based proton distributions is examined in Appendix \ref{appsec:flow-conv}. 
The results shown in Fig.~\ref{fig:conv} indicate that as the flow model is trained, the proton distribution converges toward the target Boltzmann distribution defined by the energy function in Eq.~\eqref{eq:elecwfn}.

We perform a detailed analysis of the proton configurations for $r_s$ values larger than 1.41 to gain deeper insight into the molecular phase. Panel (a) of Figure~\ref{fig:com} presents a snapshot of the sampled atomic configurations at $r_s=1.43$.
Hydrogen molecules are identified by pairing the closest hydrogen atoms, with their centers of mass marked by red spheres. Notably, these red spheres do not follow a random liquid arrangement, but instead form an ordered structure reminiscent of a solid.
To further investigate these structures, we plot the averaged simulated X-ray diffraction (XRD) spectra for an ensemble of hydrogen configurations at three different $r_s$ values in Figure~\ref{fig:com} (c). 
The delta peaks in the XRD spectra indicate that these systems are in a solid state.
We quench the atomic configurations by sampling from $p(\boldsymbol{X})^{\beta}$ with $\beta = 20$ to better characterize the structures. 
Figure ~\ref{fig:com} (b) shows the RDF of the original and quenched structures.  
As the system cools down, the features around $2 ~ a_B$ and $3 ~ a_B$ become more prominent. 
Note that in this work, we use a cubic unit cell setup as in all the liquid studies, which prevents us from examining the detailed crystal structure of the solid state. Nevertheless, our results clearly indicate the emergence of a high-temperature molecular solid at the temperature where PBE-level calculations predict a liquid-liquid phase transition.

\section{Conclusion and Discussion}
\label{sec:discussion}

In this work, we observe evidence of a direct transition from atomic liquid to molecular crystal within a temperature regime believed to be relevant to the liquid-liquid transition in the dense hydrogen system. 
This finding arises from a variational free energy calculation with many-body precision, demonstrating that the flow model is capable of transforming the structureless liquid state of the base proton distribution into a solid-state solution using many-body precision energy.
However, since the simulation is performed with a finite box containing 32 hydrogen atoms, finite-size effects may bias both the existence and the structure of the discovered solid states. 
Nevertheless, this finding aligns with recent discoveries of a possible solid-state phase in similar temperature and pressure regimes obtained through different computational methods and much larger simulation cells~\cite{niu2023stable, goswami2024high}, although with different detailed characterizations of the solid-state phase. Collectively, these works suggest that the liquid to solid transition may occur at a higher temperature than previously believed, and having precise many body energy beyond DFT is crucial for stabilizing the solid at high temperature.
Replacing the expensive wavefunction model with a cheaper yet accurate machine learning force field could allow larger scale calculations, potentially bypassing some of the limitations of this study.
Furthermore, nuclear quantum effects can be incorporated into the calculation by generalizing the proton distribution to a variational density matrix, using the neural canonical transformation approach~\cite{SciPostPhys.14.6.154, Zhang2024Neural}.

\begin{acknowledgments}
    We thank Guglielmo Mazzola, Yubo Yang, and Qi Zhang for useful discussions. We thank Guglielmo Mazzola for providing the 64 hydrogen atoms reference data shown in Appendix~\ref{appsec:single-conf}. This project is supported by the National Natural Science Foundation of China under Grants No.~12122103, No.~92270107, No.~T2225018, No.~12188101, No.~T2121001 and the Strategic Priority Research Program of Chinese Academy of Sciences under Grants No.~XDB0500000 and No.~XDB30000000.
\end{acknowledgments}

\section*{Data Availability}
The numerical data used to generate all figures in this manuscript are available in the GitHub repository \cite{data}.

\appendix
\section{Details of the neural network models}
\label{appsec:models}

\subsection{periodic features}
\label{appsec:feat}

The first step in constructing the neural networks for flow and the electron wavefunction is to introduce a periodic function $\bm f$ that transforms the coordinates of protons and electrons into a form that remains invariant under both spatial translations of \emph{all} particles by any vector $\bm s  \in \mathbb{R}^3$  and the periodic transformation of \emph{any} individual particle by the lattice vector $\bm n$, i.e. 
\begin{align}\label{eq:feat-trans}
    & \bm f(\bm \Xi) = \bm f(\bm \Xi+\bm s), \\\label{eq:feat-periodicity}
    & \bm f( \dots, \bm \xi_\alpha, \dots) = \bm f(\dots, \bm \xi_\alpha + \bm n, \dots) .
\end{align}
We use the following periodic feature in our networks
\begin{align}\label{eq:periodic-feat}
&\bm f_{\alpha\beta} (\bm \Xi) =
\mathrm{concat}\Big(
   \bm f^\mathrm{ds}_{\alpha\beta},\:
    \cos(\pi \bm \sigma_{\alpha\beta})
\Big) \, ,
\end{align}
in which $\bm \sigma_{\alpha\beta} = \sum_\mu \sigma_{\alpha\beta}^\mu \bm a_\mu$, $\sigma_{\alpha\beta}^\mu = \hat \sigma_{\alpha\beta}^\mu - 2 \lfloor (\hat \sigma_{\alpha\beta}^\mu + 1) / 2 \rfloor$ and $\hat \sigma_{\alpha\beta}^\mu = 2 \bm \xi_{\alpha\beta} \cdot \bm a^\ast_\mu$.
Here $\hat \sigma_{\alpha\beta}^\mu$ represents the direct coordinates of the relative position $\bm \xi_{\alpha\beta} = \bm \xi_\alpha - \bm \xi_\beta$, while $\sigma_{\alpha\beta}^\mu$ normalizes these coordinates to the range $[-1,1]$ by applying a shift.
The $\cos$ operator is applied component-wise to the vector $\bm \sigma_{\alpha\beta}$ in the latter part, and
the term $\bm f^{\mathrm{ds}}_{\alpha\beta}$ in this expression is the periodic features introduced by the deep-solid network~\cite{li2022ab}
\begin{align}\label{eq:periodic-feat-ds}
  &\bm f^\mathrm{ds}_{\alpha\beta} (\bm \Xi)  = 
  \mathrm{concat}\Big(
      \frac 1{2\pi}\sum_{\mu=1}^3 g( 
      \sigma_{\alpha\beta}^\mu) \bm a_\mu, \:
      d(\bm \sigma_{\alpha\beta})
  \Big) \, ,
\end{align}
in which
\begin{align}\label{eq:def-d}
  &d(\bm \sigma)^2 = \frac{1}{4\pi^2}\sum_{\mu\nu} M_{\mu\nu}(\bm \sigma)\: \bm a_\mu \cdot \bm a_\nu \, , \\
  &M_{\mu\nu}(\bm \sigma) = f^2( \sigma^{\mu})\delta_{\mu\nu} + g(\sigma^{\mu})g(\sigma^{\nu})(1-\delta_{\mu\nu}) \, ,
  \\
  &f(\sigma) = \vert\sigma\vert \Big( 1 - \frac{\vert \sigma\vert^3}{4} \Big) \, , \quad
  g(\sigma) = \sigma \Big( 1 - \frac32 \vert \sigma\vert + \frac12 \vert \sigma\vert^2 \Big) \, .
\end{align}
It is straightforward to verify that the features constructed via Eq.~\eqref{eq:periodic-feat} satisfy the conditions specified in Eqs.~\eqref{eq:feat-trans} and \eqref{eq:feat-periodicity}.

To enhance the feature representation, we introduce a multi-scale transformation for the deep-solid features,
\begin{align}
    \mathrm{multi\textunderscore scale}(\,\bm f^\mathrm{ds}_{\alpha\beta}\,)
    =
    \mathrm{concat} \Big(\{\, s_i  \cdot \bm f^\mathrm{ds}_{\alpha\beta} \,\}_{i\in S} \Big) \, ,
\end{align}
where \(S\) is a set of all possible scales.
For instance, when considering 16 protons, \(S = \{1, 2, 4\}\).
The multi-scale transformation multiplies the deep-solid feature by each scale in the set \(S\) and concatenates the results.
The multi-scale transformed feature reads
\begin{align}\label{eq:periodic-feat-ms}
&\bm f^{\mathrm{ms}}_{\alpha\beta} (\bm \Xi) =
\mathrm{concat}\Big(
    \mathrm{multi\textunderscore scale}(\bm f^\mathrm{ds}_{\alpha\beta}),\:
    \cos(\pi \bm \sigma_{\alpha\beta})
\Big).
\end{align}
This feature is employed by default throughout this work unless stated otherwise.

\subsection{Flow network}
\label{appsec:flow}

{\linespread{1.5}
\begin{algorithm} [H]
    \caption{Flow ans\"atz}
    \label{alg:flow}
    \begin{algorithmic}[1]
      \Require
      Proton coordinates $\bm x_I$.
      \State
      $\bm x_{IJ} \gets \bm x_{I} - \bm x_{J}$
      \State
      $\bm f^\mathrm{ms}_{IJ} \gets \textrm{features} (\bm x_{IJ})$
      \algorithmiccomment{Eq.~\eqref{eq:periodic-feat-ms}}
      \State
      $\bm h^{(0)}_{I} \gets {\mathrm{linear}}\left(\frac1N \sum_{J}\bm f^\mathrm{ms}_{IJ} \right)
      \quad
      \bm h^{(0)}_{IJ} \gets {\mathrm{linear}}\left(\bm f^\mathrm{ms}_{IJ} \right)
      $
      \For{$i = 0$ to $L-1$}
      \State
      $\bm g^{(l)}_I \gets \textrm{concat}\left(\bm h^{(l)}_I, \frac{1}{N}\sum_I\bm h^{(l)}_I, \frac 1N\sum_J \bm h^{(l)}_{IJ}\hat{\bm h}^{(l)}_J\right)$
      \State
      $\bm h^{(l+1)}_I\gets \frac{1}{\sqrt 2} \bm h^{(l)}_I + \frac{1}{\sqrt 2} \textrm{dense\textunderscore layer} \big(\bm g^{(l)}_I\big)$
      \State
      $\bm h^{(l+1)}_{IJ} \gets \frac{1}{\sqrt 2} \bm h^{(l)}_{IJ} + \frac{1}{\sqrt 2} \textrm{dense\textunderscore layer} \big(\bm h^{(l)}_{IJ}\big)$
      \EndFor
      \State
      $\bm h_I(\bm X) \gets \bm h^{(L)}_I$
      \State
      $\bm{Z}_I(\bm X) \gets \bm{X}_I + {\text{linear}} \, (\bm{h}_I \big(\bm X)\big)$ 
      \\
      \Return  $\bm Z_I(\bm X)$
    \end{algorithmic}
\end{algorithm}}

In Algorithm~\ref{alg:flow}, we introduce single-particle features $\bm h_I$ and pair-particle features $\bm h_{IJ}$ with dimensions ${d^p_1}$ and ${d^p_2}$.
These features are initialized using a multi-scale periodic feature function (line 2) and trainable linear mappings (line 3) which reshape their feature dimensions to ${d^p_1}$ and ${d^p_2}$.
The features are then subsequently updated through $L$ message-passing layers, with the update at the $(l+1)$-th layer described by lines 5-7.
The hat over $\bm h^{(l)}_J$ indicates that the single-particle feature is linearly projected to the dimension $d^p_2$, so the multiplication is performed component-wise.
The ``dense\textunderscore layer'' operation consists of a trainable linear mapping followed by an element-wise hyperbolic tangent activation.
Finally, a trainable linear mapping is applied to the feature dimension $d^p_1$ to produce a 3-dimensional vector, and the coordinates are updated via a backflow architecture (line 10). 

\subsection{Electron wavefunction}
\label{appsec:elecwfn}

{\linespread{1.5}
\begin{algorithm} [H]
    \caption{Wavefunction ans\"atz}
    \label{alg:model}
    \begin{algorithmic}[1] 
      \Require
      Coordinates $\bm \xi_\alpha$, one-hot $\bm t_\alpha$, kpoints $\bm k_m$, twist $\bm w$, scale $r_s$.
      \State
      $\bm \phi_i^\delta, \bm h_\alpha \gets \textrm{orbitals}(\bm \xi_{\alpha}, \bm t_{\alpha}, \bm w, r_s)$
      \algorithmiccomment{Algorithm~\ref{alg:orbitals}}
      \State
      $ \bm e_i^\delta \gets \textrm{envelope} (\bm h_I, \bm r_{iI}, r_s) $
      \algorithmiccomment{Eq.~\eqref{eq:orb-0} and \eqref{eq:orb-1}}
      \State
      $ \bm \phi_i^\delta \gets \bm e_i^\delta \odot \bm \phi_i^\delta$
      \State
      $D^\delta_{ij} \gets  
      \sum_\mu W_{\mu} \phi^\delta_{i,\mu}\phi^\delta_{j,\mu} $
      \algorithmiccomment{$i\in\uparrow$, $j\in\downarrow$}
      \State
      $\lambda_m^\delta \gets \textrm{MLP} (\bm w, r_s) $
      \State
      $ \hat {\bm r}_i \gets \bm r_i + \textrm{linear}(\bm h_i) $
      \State
      $ E^\delta_{ij} \gets \frac 1 {\sqrt{n_kV}} \sum_{m=1}^{n_k} \lambda_m^\delta e^{ 2\pi i (\bm k_m + \bm w) \cdot (\hat {\bm r}_i - \hat {\bm r}_j) } $
      \algorithmiccomment{$i\in\uparrow$, $j\in\downarrow$}
      \State
      $ \Psi \gets \sum_{\delta} \det ( E^\delta \odot D^\delta )  $       
      \algorithmiccomment{$\odot$ element-wise multiplication}
      \\
      \Return  $\Psi$ 
    \end{algorithmic}
\end{algorithm}}

The wavefunction ansatz is outlined in Algorithm~\ref{alg:model}.
In this algorithm, electrons and protons are treated equivalently and are distinguished by a one-hot type coding \(\bm{t}_\alpha\), i.e. electrons in the \(\uparrow\) and \(\downarrow\) channels are encoded as \((1, 0, 0)\) and \((0, 1, 0)\), respectively, while protons are encoded as \((0, 0, 1)\).
This type coding can be extended to systems with heavier nucleis.
In addition to the coordinates and type coding, we also input the k-space sampling points \(\bm{k}_m\), the twist \(\bm{w}\), and the scale \(r_s\) as parameters to the wavefunction.
With these inputs, the coordinates are first processed to construct the feature function $\bm{h}_\alpha$, the electron orbital $\bm{\phi}_i^\delta$, and the electron envelope $\bm{e}_i^\delta$ (lines 1-2).
The geminal matrix~\cite{casula2003geminal} $D_{ij}^\delta$ is computed as a bilinear form of the spin-up and spin-down electron orbitals, with a diagonal trainable parameter matrix $W_\mu$ (line 4).
The geminal envelope $E_{ij}^\delta$ is constructed as a linear combination of plane waves with twist boundary conditions and a backflow transformation of the electron coordinates (lines 5-7).
Finally, the wavefunction is expressed as a linear combination of the determinants of the geminals (line 10).

{\linespread{1.5}
\begin{algorithm}[H]
    \caption{Electron orbitals}
    \label{alg:orbitals}
    \begin{algorithmic}[1]
      \Require{
        Coordinates $\bm \xi_{\alpha}$,
        one-hot $\bm t_\alpha$,
        twist $\bm w$ and
        scale $r_s$. 
      }
      \State
      $\bm \xi_{\alpha\beta} \gets \bm \xi_{\alpha} - \bm \xi_{\beta}$
      \State
      $\bm f^\mathrm{ms}_{\alpha\beta} \gets \textrm{features} (\bm \xi_{\alpha\beta})$
      \algorithmiccomment{Eq.~\eqref{eq:periodic-feat-ms}}
        \State
        $\bm v \gets \mathrm{concat}(\bm w, r_s)$,
        \State
        $\bm h^{(0)}_\alpha \gets \mathrm{linear}\Big( \mathrm{concat}\Big( \frac{1}{n} \sum_\beta \bm f^\mathrm{ms}_{\alpha\beta}, \, \bm t_\alpha, \bm v \Big)\Big)$, 
        \State
        $\bm h^{(0)}_{\alpha\beta} \gets \mathrm{linear}\Big(\mathrm{concat}\Big( \bm f^\mathrm{ms}_{\alpha\beta}, \bm t_\alpha, \bm t_\beta, \bm v \Big)\Big) $, 
        \For{$l = 0$ to $L-1$}
            \State $\bm g^{(l)}_\alpha \gets \mathrm{intermediate\textunderscore feature}(\bm h^{(l)}_\alpha)$
            \algorithmiccomment{see Eq.~\eqref{eq:inter-feat-0} and \eqref{eq:inter-feat-1}}
            \State
            $\bm h^{(l+1)}_\alpha \gets \frac{1}{\sqrt 2} \bm h^{(l)}_\alpha + \frac{1}{\sqrt 2} \textrm{dense\textunderscore layer} \big(\bm g^{(l)}_\alpha\big) $
            \State
            $\bm h^{(l+1)}_{\alpha\beta} \gets \frac{1}{\sqrt 2} \bm h^{(l)}_{\alpha\beta} + \frac{1}{\sqrt 2} \textrm{dense\textunderscore layer} \big(\bm h^{(l)}_{\alpha\beta}\big) $
            \If {do\textunderscore attn}
            \State
            $\bm h^{(l+1)}_{\alpha\beta} \gets \frac{1}{\sqrt 2} \bm h^{(l)}_{\alpha\beta} + \frac{1}{\sqrt 2} \textrm{gated\textunderscore atten\textunderscore layer} \big(\bm h^{(l)}_{\alpha\beta}\big) $
            \EndIf
        \EndFor
        \State
        $\bm \phi^\delta_i \gets \textrm{linear}(\bm h_i^{(L)}) $ \algorithmiccomment{$\delta$ is the index of determinants}\\
        \Return  $\bm \phi^\delta_i$, $\bm h_\alpha^{(L)}$ 
    \end{algorithmic}
\end{algorithm}}

The electron orbitals are constructed in a Ferminet-like manner, with single- and pair-particle features defined symmetrically for both electrons and protons.
In each layer, an intermediate feature $\bm g$ is constructed as follows
\begin{align}\nonumber
  \bm g^{(l)}_I = \textrm{concat}
  \bigg(&
          \bm h^{(l)}_I,
          \frac{1}{N}\sum_I\bm h^{(l)}_I,
          \frac{1}{N}\sum_I\bm h^{(l)}_I,
          \frac{1}{n^\uparrow}\sum_{i\in\uparrow} \bm h^{(l)}_{Ii} \hat{\bm h}^{(l)}_i,\\\label{eq:inter-feat-0}
        & \frac{1}{n^\downarrow}\sum_{i\in\downarrow}  \bm h^{(l)}_{Ii} \hat{\bm h}^{(l)}_i,
          \frac{1}{N}\sum_{J} \bm h^{(l)}_{IJ} \hat{\bm h}^{(l)}_J
          \bigg), \\ \nonumber
  \bm g^{(l)}_i = \textrm{concat}
  \bigg(&
          \bm h^{(l)}_i,
          \frac{1}{n^\uparrow}\sum_i\bm h^{(l)}_i,
          \frac{1}{n^\downarrow}\sum_i\bm h^{(l)}_i,
          \frac{1}{n^\uparrow}\sum_{j\in\uparrow}  \bm h^{(l)}_{ij} \hat{\bm h}^{(l)}_j,\\\label{eq:inter-feat-1}
        & \frac{1}{n^\downarrow}\sum_{j\in\downarrow}  \bm h^{(l)}_{ij} \hat{\bm h}^{(l)}_j,
          \frac{1}{N}\sum_I  \bm h^{(l)}_{iI}\hat{\bm h}^{(l)}_I
          \bigg).
\end{align}
The hat over \(\bm{h}^{(l)}_\alpha\) indicates that the single-particle feature is linearly projected to the same dimension as \(\bm{h}_{\alpha\beta}\), enabling component-wise multiplication.

We introduce a gated attention layer during the message-passing procedure to enhance the network's generalization ability. 
The gated self-attention Algorithm~\ref{alg:attn} is like the standard self-attention algorithm, except the gate term inspired by the AlphaFold~\cite{jumper2021highly}. 
The query, key and value are the linear mappings from the pair-particle channel. 
The attention map is formed by the softmax applied to the inner product of queries and keys connecting to the same particle. 
The gate is applied as a component-wise multiplication on the attention mapped values. 
Finally, the values in different heads are concatenated and linearly mapped to give the output pair-particle channel. 

{\linespread{1.5}
\begin{algorithm}[H]
    \caption{Gated self-attention}
    \label{alg:attn}
    \begin{algorithmic}[1]
      \Require{Pair-particle channel $\bm h_{\alpha\beta}$}
        \State
        $\bm h_{\alpha\beta} \gets \textrm{layer\textunderscore norm} (\bm h_{\alpha\beta})$
        \State
        $\bm q^h_{\alpha\beta}, \bm k^h_{\alpha\beta}, \bm v^h_{\alpha\beta} \gets \textrm{linear\textunderscore no\textunderscore bias}(\bm h_{\alpha\beta})$
        \State
        $\bm g^h_{\alpha\beta} \gets \mathrm{sigmoid}\left(\textrm{linear}(\bm h_{\alpha\beta})\right)$
        \State
        $m^h_{\alpha\gamma\beta} \gets \textrm{softmax}_{\gamma}( \bm q^h_{\alpha\beta}{}^{\top} \bm k^h_{\gamma\beta} / \sqrt{d_h} )  $ 
        \State
        $\bm u^h_{\alpha\beta} \gets \bm g^h_{\alpha\beta}\odot \sum_\gamma m^h_{\alpha\gamma\beta}\bm v^h_{\gamma\beta}$
        \algorithmiccomment{$\odot$ element-wise multiplication}
        \State
        $ \hat{\bm h}_{\alpha\beta} \gets \textrm{linear\textunderscore no\textunderscore bias}( \textrm{concat}_h (\bm u^h_{\alpha\beta}) )  $\\
        \Return  $\hat{\bm h}_{\alpha\beta}$ 
    \end{algorithmic}
\end{algorithm}
}
\begin{figure}[tbh]
  \centering
  \includegraphics[scale=1]{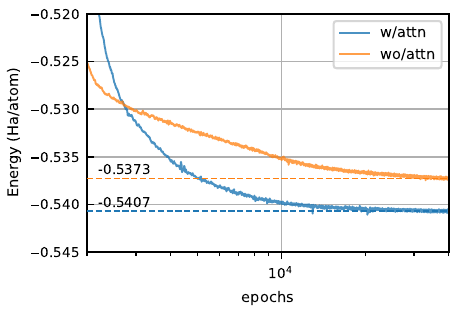}
  \caption{Comparison of ensemble training curves with and without the attention layer.}
  \label{fig:attn}
\end{figure}

Figure~\ref{fig:attn} shows a comparison of the ensemble {VMC} training results for systems with 32 hydrogen atoms, both with and without the attention layer.
The training curve clearly shows that the attention layer significantly enhances the expressibility of the wavefunction in the multi-configuration case, leading to a lower variational energy.

The orbital envelope is given by:
\begin{align}\label{eq:orb-0}
  &\bm e^\delta_i = \sum_I \bm\pi^\delta_I \odot \exp( -d_{iI} \bm\Sigma^\delta_I  ) \\\label{eq:orb-1}
  &  \bm \pi^\delta_I= 1.0 + \textrm{MLP}(\bm h_I), \quad \bm \Sigma^\delta_I = r_s + \textrm{MLP}(\bm h_I),
\end{align}
where the periodic distance \(d_{iI} = d(\bm{\sigma}_{iI})\) (defined by Eq.~\eqref{eq:def-d}) between electron \(i\) and proton \(I\) is scaled by the trainable parameters \(\bm{\Sigma}^\delta_I\) to provide the input for the exponent.
The exponent is then scaled by the trainable parameters \(\bm{\pi}^\delta_I\) to form the envelope of proton \(I\).
The envelopes are summed over all protons to provide the contribution of the \(i\)-th orbital of determinant \(\delta\).
Notably, all electrons share the same parameters \(\bm{\pi}^\delta_I\) and \(\bm{\Sigma}^\delta_I\).
The envelope is isotropic, meaning that the decay with respect to the distance \(d\) is uniform across all three spatial dimensions.

\section{Benchmarks of the single configuration variational calculations}
\label{appsec:single-conf}

\begin{table}[bht]
  \centering
\caption{Comparisons of energies of 64H atoms at different temperatures and $r_s$ values.}
\label{tab:energy-64H}
\begin{tabular}{@{}ccccc}
  \multicolumn{5}{@{}r@{}}{\footnotesize{Ha/atom}} \\
  \toprule
  &  & Ref.~\onlinecite{PhysRevLett.120.025701} & This work & \\
  \midrule
  & 1200K\phantom{12} &&& \\ 
  \midrule
  &$r_s = 1.34$ & \phantom{12}-0.52106(4) & \phantom{12}-0.52311(5) & \\
  \midrule
  &$r_s = 1.42$ & \phantom{12}-0.54047(4) & \phantom{12}-0.54257(5) & \\
  \midrule
  \midrule
  & 6000K\phantom{12} &&& \\ 
  \midrule
  &$r_s = 1.82$ & \phantom{12}-0.55229(4) & \phantom{12}-0.55417(4) & \\
  \midrule
  &$r_s = 1.94$ & \phantom{12}-0.56591(4) & \phantom{12}-0.56801(3) & \\
  \bottomrule
\end{tabular}
\end{table}

To demonstrate the accuracy of our neural-network wavefunction for single-configuration training, we benchmark it against the results computed using the setup in Ref.~\onlinecite{PhysRevLett.120.025701} for 64 hydrogen atoms under periodic boundary conditions at different densities, as shown in Table \ref{tab:energy-64H}.  
At $r_s = 1.34$ and 1200K, as well as $r_s = 1.82$ and 6000K, the system is in the atomic phase; at $r_s = 1.42$ and 1200K, as well as $r_s = 1.94$ and 6000K, the system is in the molecular phase.
The results clearly shows that the neural network wavefunction employed in this paper consistently achieves lower variational energies across all test cases.

\section{The DeePMD model}  
\label{appsec:deepmd}

The DeePMD model used in this work is developed via the concurrent learning algorithm implemented in the DP-GEN package \cite{dpgen}.
The model's quality is iteratively enhanced by expanding the training dataset with data that are essential for improving its accuracy in this framework.
Each iteration of concurrent learning involves three phases: training, exploration, and DFT calculation.

DFT-calculated energies, forces, and virial tensors serve as labels for training the DeePMD model.
The calculations in this work are performed using the ABACUS package \cite{abacus1,abacus2}, employing the Perdew-Burke-Ernzerhof (PBE) generalized gradient approximation \cite{Perdew1996PBE} for the exchange-correlation functional and a plane-wave basis.
The plane-wave energy cutoff is set to 100~Ry, and the k-space is sampled with a grid spacing of 0.15~bohr\(^{-1}\).

Starting with an initial dataset generated from short ab initio molecular dynamics (AIMD) simulations of known hydrogen phases I, II, and III, we trained an ensemble of four DeePMD models using the smooth edition of the DeePMD descriptor~\cite{end-to-end}.
Each model employs a cut-off radius of 6 \AA, an embedding network with neuron counts of [25, 50, 100], and a fitting network with neuron counts of [240, 240, 240].
All models share the same training data and hyperparameters but use different random seeds for parameter initialization. Each model is trained for 100,000 steps.

During the exploration phase, isothermal-isobaric (NPT) MD simulations are performed by sampling configurations within a temperature range of 50-3000 K and a pressure range of 10-300 GPa.
In the early iterations, simulations primarily focus on relatively low temperatures and pressures, while the later iterations extend to higher temperatures and pressures. 
Throughout the molecular dynamics simulation trajectories, the maximum standard deviation of the atomic forces—referred to as the model deviation—is used to estimate force-prediction errors. 
Frames exhibiting a model deviation exceeding 0.5 times the mean absolute magnitude of the predicted forces are flagged as significantly erroneous and potentially indicative of non-physical configurations, such as atomic overlaps.
Up to 300 frames are randomly sampled from those exhibiting model deviation values within a candidate range, defined as 0.2 to 0.5 times the mean absolute magnitude of the forces.
The energies, forces, and virial tensors associated with these configurations are computed using DFT and incorporated into the training dataset for subsequent iterations.
This exploration process gradually enriches the dataset with liquid configurations in addition to the solid ones, enhancing the model's ability to describe both phases.

As the iterations proceed, the model's quality gradually improves, reducing the number of configurations whose model deviations are within the candidate range. Once the number of such configurations no longer decreases, we consider the concurrent learning process convergent.
In total, 84 concurrent learning iterations were conducted, resulting in 28,537 labeled configurations in the training dataset.
A final production DeePMD model was then trained on this dataset for 16,000,000 steps, achieving training accuracies of 0.012~eV/atom for energy, 0.276~eV/\AA\ for forces, and 0.021~eV/atom for the virial tensor. The trained model and the input files can be found in the Github repository \href{https://github.com/fermiflow/SolidHydrogen}{https://github.com/fermiflow/SolidHydrogen}.

\section{Convergence of the proton distribution}
\label{appsec:flow-conv}

\begin{figure}[tbh]
  \centering
  \includegraphics[width=\columnwidth]{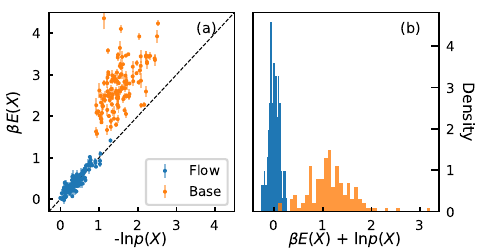}
  \caption{Comparison between the proton distribution sampled from the flow model and the target distribution defined by the electronic wavefunction of 32H at $T = 1200$ K and $r_s = 1.44$ (with the minimum value of $-\ln p(\boldsymbol{X})$ shifted to zero).
  (a) Scatter plot comparing the model density and the target density.
  (b) Histogram of the differences between the two distributions.}
  \label{fig:conv}
\end{figure}

To inspect the convergence of the proton distribution parameterized by the flow model (see Eq.~\ref{eq:flow}), we follow the analysis in Ref.~\onlinecite{Wirnsberger_2022}. 
Given an energy function $E(\mathbf{X})$, the target proton distribution is defined as $e^{-\beta E(\mathbf{X})} / \mathcal{Z}$, where $\beta = 1/k_B T$ and $\mathcal{Z}$ is the partition function.
At convergence, the flow-based distribution satisfies $p(\mathbf{X}) \propto e^{-\beta E(\mathbf{X})}$, which is equivalent to
\begin{align}
  \text{ln} p(\mathbf{X}) = - \beta E(\mathbf{X}) + \text{constant} \, ,
\end{align}
with the constant related to the unknown partition function.
Figure \ref{fig:conv} compares distributions computed for the trained flow model and the base distribution, with the constant shift estimated by the average difference between $\beta E(\mathbf{X})$ and $- \text{ln} p(\mathbf{X})$ of the trained flow model.
In our calculations, the energy $E(\mathbf{X})$ is evaluated using a trial wavefunction similar to the VMC method, and is therefore subject to statistical noise. We sampled 128 proton configurations for each distribution and generated 3200 electron configurations for each proton configuration to estimate the energy.
From the scatter plot shown in panel (a), it is evident that with the trained flow model, most samples lie around the reference line with slope one, whereas the samples from the base distribution exhibit significantly larger deviations and do not show any linear relation.
The histograms in panel (b) further demonstrate that the distribution of the trained model shifts toward lower values as a result of the variational optimization of the flow model. Moreover, the distribution becomes more sharply peaked, reflecting a reduction in the variance of the variational free energy.

\bibliography{ref}

\end{document}